\documentclass[a4paper]{elsarticle}

\usepackage{lineno,hyperref}
\usepackage{subfig}
\usepackage{multirow}
\modulolinenumbers[5]
\usepackage{geometry} 
\usepackage{multicol}
\usepackage{tabularx,ragged2e,booktabs,caption}
\biboptions{numbers,sort&compress}

\journal{Physics Letters B}

\begin{document}

\begin{frontmatter}

\title{Double-crystal setup measurements at the CERN SPS}
%\tnotetext[mytitlenote]{Fully documented templates are available in the elsarticle package on \href{http://www.ctan.org/tex-archive/macros/latex/contrib/elsarticle}{CTAN}.}

%% Group authors per affiliation:
%\author{Elsevier\fnref{myfootnote}}
%\address{Radarweg 29, Amsterdam}
%\fntext[myfootnote]{Since 1880.}

%% or include affiliations in footnotes:
%\author[mymainaddress,mysecondaryaddress]{Elsevier Inc}
%\ead[url]{www.elsevier.com}

%\author[mysecondaryaddress]{Global Customer Service\corref{mycorrespondingauthor}}
%\cortext[mycorrespondingauthor]{Corresponding author}
%\ead{support@elsevier.com}

\author[CERN_address]{W.~Scandale}
\author[CERN_address]{F.~Cerutti}
\author[CERN_address]{L.S.~Esposito}
\author[CERN_address,ICL_address]{M.~Garattini}
\author[CERN_address]{S.~Gilardoni}
\author[CERN_address]{S.~Montesano}
\author[CERN_address]{R.~Rossi}
\author[PSUD_address]{L.~Burmistrov}
\author[PSUD_address]{S.~Dubos}
\author[CERN_address,PSUD_address,TSUNK_address]{A.~Natochii\corref{correspondingauthor}}
\cortext[correspondingauthor]{Corresponding author}
\ead{andrii.natochii@cern.ch}
\author[PSUD_address]{V.~Puill}
\author[PSUD_address]{A.~Stocchi}
\author[CERN_address,TSUNK_address]{V.~Zhovkovska}
\author[CERN_address,INFN_address1]{F.~Murtas}
\author[INFN_address2]{F.~Addesa}
\author[INFN_address2]{F.~Iacoangeli}
\author[INFN_address3]{F.~Galluccio}
\author[JINR_address]{A.D.~Kovalenko}
\author[JINR_address]{A.M.~Taratin}
\author[CERN_address,JINR_address]{G.I.~Smirnov}
\author[PNPI_address]{A.S.~Denisov}
\author[PNPI_address]{Yu.A.~Gavrikov} 
\author[PNPI_address]{Yu.M.~Ivanov}
\author[PNPI_address]{L.P.~Lapina} 
\author[PNPI_address]{L.G.~Malyarenko}
\author[PNPI_address]{V.V.~Skorobogatov}
\author[IHEP_address]{A.G.~Afonin}
\author[IHEP_address]{Yu.A.~Chesnokov}
\author[IHEP_address]{A.A.~Durum}
\author[IHEP_address]{V.A.~Maisheev}
\author[IHEP_address]{Yu.E.~Sandomirskiy}
\author[IHEP_address]{A.A.~Yanovich}
\author[ICL_address]{J.~Borg}
\author[ICL_address]{T.~James}
\author[ICL_address]{G.~Hall}
\author[ICL_address]{M.~Pesaresi}

\address[CERN_address]{The European Organization for Nuclear Research (CERN), CH-1211 Geneva 23, Switzerland}
\address[ICL_address]{Imperial College, London, The United Kingdom}
\address[PSUD_address]{Laboratoire de l'Acc\'el\'erateur Lin\'eaire (LAL), Universit\'e Paris-Sud, Orsay, France}
\address[TSUNK_address]{On leave from Taras Shevchenko National University of Kyiv (TSNUK), 60 Volodymyrska Street, 01033 Kyiv, Ukraine}
\address[INFN_address1]{INFN, Laboratori Nazionali di Frascati, Via Fermi, 40 00044 Frascati (Roma), Italy}
\address[INFN_address2]{INFN Sezione di Roma, Piazzale Aldo Moro 2, 00185 Rome, Italy} 
\address[INFN_address3]{INFN Sezione di Napoli, Complesso Universitario di Monte Sant'Angelo, Via Cintia, 80126 Napoli, Italy}
\address[JINR_address]{Joint Institute for Nuclear Research (JINR), Joliot-Curie 6, 141980 Dubna, Russia}
\address[PNPI_address]{Petersburg Nuclear Physics Institute in National Research Centre "Kurchatov Institute" (PNPI), 188300 Gatchina, Russia}
\address[IHEP_address]{Institute for High Energy Physics (IHEP), National Research Centre Kurchatov Institute, Protvino, Moscow region, 142281 Russia}

\begin{abstract}
In this paper, we discuss an experimental layout for the two-crystals scenario at the Super Proton Synchrotron (SPS) accelerator. The research focuses on a fixed target setup at the circulating machine in a frame of the Physics Beyond Colliders (PBC) project at CERN. The UA9 experiment at the SPS serves as a testbench for the proof of concept, which is planning to be projected onto the Large Hadron Collider (LHC) scale. The presented in the text configuration was used for the quantitative characterization of the deflected particle beam by a pair of bent silicon crystals. For the first time in the double-crystal configuration, a particle deflection efficiency by the second crystal of $0.188 \pm 3 \cdot 10^{-5}$ and $0.179 \pm 0.013$ was measured on the accelerator by means of the Timepix detector and Beam Loss Monitor (BLM) respectively. In this setup, a wide range angular scan allowed a possibility to \textit{in situ} investigate different crystal working regimes (channeling, volume reflection, etc.), and to measure a bent crystal torsion.
\end{abstract}
\begin{keyword}
Channeling, Double-Crystal, Accelerator, Timepix, Beam Loss Monitor
%\MSC[2010] 00-01\sep  99-00
\end{keyword}

\end{frontmatter}

%\linenumbers

\begin{multicols}{2}

\section{Introduction}

In September 2016 a new study of the opportunities offered by the CERN's accelerator complex and its scientific infrastructure has been launched. The aim of the PBC project to focus on fundamental physics inquiries that are similar in substance to those referred to high-energy colliders, but that need different types of experiments and beams~\cite{DOUBLECHANNELING1}. It will explore CERN's opportunities for a vision of some opened questions in high energy and particle physics.

In a frame of the PBC, and to collect different physical proposal for the LHC, a working group on fixed target physics has been created in 2017. The main idea of the most offers of the fixed target experiments is to re-use (probably, with small modifications) the existing detectors installed at the collider, without significant perturbation of the working conditions (in terms of luminosity, background, beam time, etc.) for the presently operating experiments (ATLAS, CMS, ALICE, LHCb, etc.).

At the moment, there are three main proposals for the implementation of the fixed target experiment~\cite{DOUBLECHANNELING2}: (1) beam halo extraction, based on a bent crystal; (2) the use of unpolarized targets; (3) the use of polarized targets.

In this paper, we will focus only on the crystal-based beam splitting technique. A description of others can be found in~\cite{DOUBLECHANNELING2}. The approach utilizes a bent crystal, used for primary LHC beam collimation studies~\cite{EXTRA7}, to direct halo particles onto a target installed in front of a particle identification detector (PID), such as LHCb~\cite{EXTRA5,DOUBLECHANNELING3}. Another bent crystal with a higher deflection angle may be installed after the target to measure the magnetic and electric dipole moments of short-living baryons produced in the target, as proposed in~\cite{EXTRA1,EXTRA2,EXTRA3,EXTRA4}. 

Such a crystals configuration is called a double-crystal setup, and before to be implemented at the LHC, it should be verified at the SPS machine. Therefore, in a frame of the UA9 collaboration~\cite{EXTRA6}, which has an experimental section at the SPS accelerator, these measurements were planned for 2017 and 2018~\cite{DOUBLECHANNELING3}.

\section{UA9 Collaboration and experimental layout}

The planar channeling appears when a charged relativistic particle enters the orientated crystal within a certain angular range close to the atomic planes. The angular acceptance for the process (critical angle $\pm\theta_{\rm c}$) depends on the particle momentum and on the interplanar potential well of the crystal: $\theta_{\rm c} = \sqrt{2U_{\rm max}/pv}$, where $U_{\rm max}$ is a potential well ($\rm\sim20~eV$ for Si) between two neighbour crystalline planes, $p$ and $v$ are the particle momentum and velocity respectively. A bent crystal deflects the particles by hundreds or thousands of microradians. The detailed description of the particle interaction processes inside a crystalline lattice can be found in~\cite{CHANNELING_2}. 

Bent crystals are becoming widely used in HEP applications, in particular for accelerator physics. The first results for the crystal extraction of the high-energy beams were obtained at CERN~\cite{UA9_83,UA9_84,UA9_1}, IHEP-Protvino~\cite{UA9_2}, RHIC~\cite{UA9_3} and FNAL~\cite{UA9_4}.

In 2008 the UA9 experiment~\cite{EXTRA6} was approved by CERN for the investigation of the crystal-based collimation technique for high-energy proton and ion beams for the LHC~\cite{UA9_5}. The main goal of the experiment is to show, how much a crystal-based collimation system is more efficient with respect to the traditional one~\cite{UA9_6}.

Since 2009 the UA9 experiment uses the SPS accelerator for investigation of the channeling effect in a bent crystal and development of the crystal-based collimation technique for the LHC. Multiturn channeling measurements (each particle in the halo can pass the crystal several time along its circular movement) are carried out in the SPS Long Straight Section~5 (LSS5)~\cite{UA9_12}, and for a single-pass channeling (the particle crosses a crystal only once along its path) efficiency investigation it was placed at the H8 beamline of the CERN North Area. The beamline is equipped by a tracker, which consists of five position sensitive microstrip planes, with an angular resolution of about $\rm 5~\mu rad$ for a 400~GeV/c proton beam~\cite{UA9_13,UA9_80,UA9_81}.

Detailed information about the SPS UA9 components alignment procedure is well described in~\cite{UA9_71}.The layout consists of two IHEP goniometers (with a couple of bent silicon crystals per goniometer). Each goniometer provides a horizontal linear movement and rotation of crystals with a designed resolution of $\rm 1~\mu rad$ and an accuracy close to $\rm 10~\mu rad$. The deflected beam is intercepted by a movable tungsten absorber (TAL). Its cross section is $\rm 70\times60~mm^{2}$ and length of 60~cm along the beam, placed 60~m downstream the crystals with a phase advance of about 60$^\circ$. In the region between crystals and TAL an LHC-type collimator with two horizontal graphite (1~m of length) jaws is located. About 1.5~m downstream the collimator there is a Roman Pot (RP) tank with a Timepix detector~\cite{UA9_36} installed inside a secondary vacuum. A horizontal motorized linear axis is used for the detector alignment. The Pot has a 0.2~mm wide aluminium wall which is 3.4~cm thick along the beam direction. There are different types of detectors installed close to each device, outside the beam pipe, such as polystirene scintillators and LHC-type BLMs~\cite{EXTRA8,UA9_12,UA9_71}, which are composed of a 50~cm long cylinder of 9~cm diameter, filled with $\rm N_{2}$ gas at 100~mbar overpressure and using an integration time of 1.2~s.

\section{The first alignment attempt}

The aim of the UA9 experiment at the SPS, in a frame of the PBC program, is to demonstrate the feasibility of the double-crystal setup at the circulating machine and its characterization in terms of the particle extraction efficiency and background production.

In September and October 2017 two dedicated runs of 24 hours each were performed for the double-crystal setup investigation. The results of the first attempt of the alignment at the circulating machine by means of the beam losses counters during a linear scan of the collimator jaw are demonstrated in~\cite{DOUBLECHANNELING4,EXTRA9}. 

A very low sensitivity of the BLM did not give a possibility to align the downstream crystal in an optimal configuration for the double-crystal setup. However, precise measurements of the beam profile and crystals alignment were provided by the Timepix silicon quantum imager. Figure~\ref{fig:fig5} illustrates an integrated image of the Timepix detector for a double-channeled beam spot, where a single-deflected beam by an upstream Crystal1 is intercepted by almost 50~m downstream Crystal2 in a channeling orientation. The shadow of the Crystal2, inside a single-deflected beam spot, serves for the crystal positioning alignment.

\begin{minipage}{0.9\linewidth}
    \centering\includegraphics[width=1\linewidth]{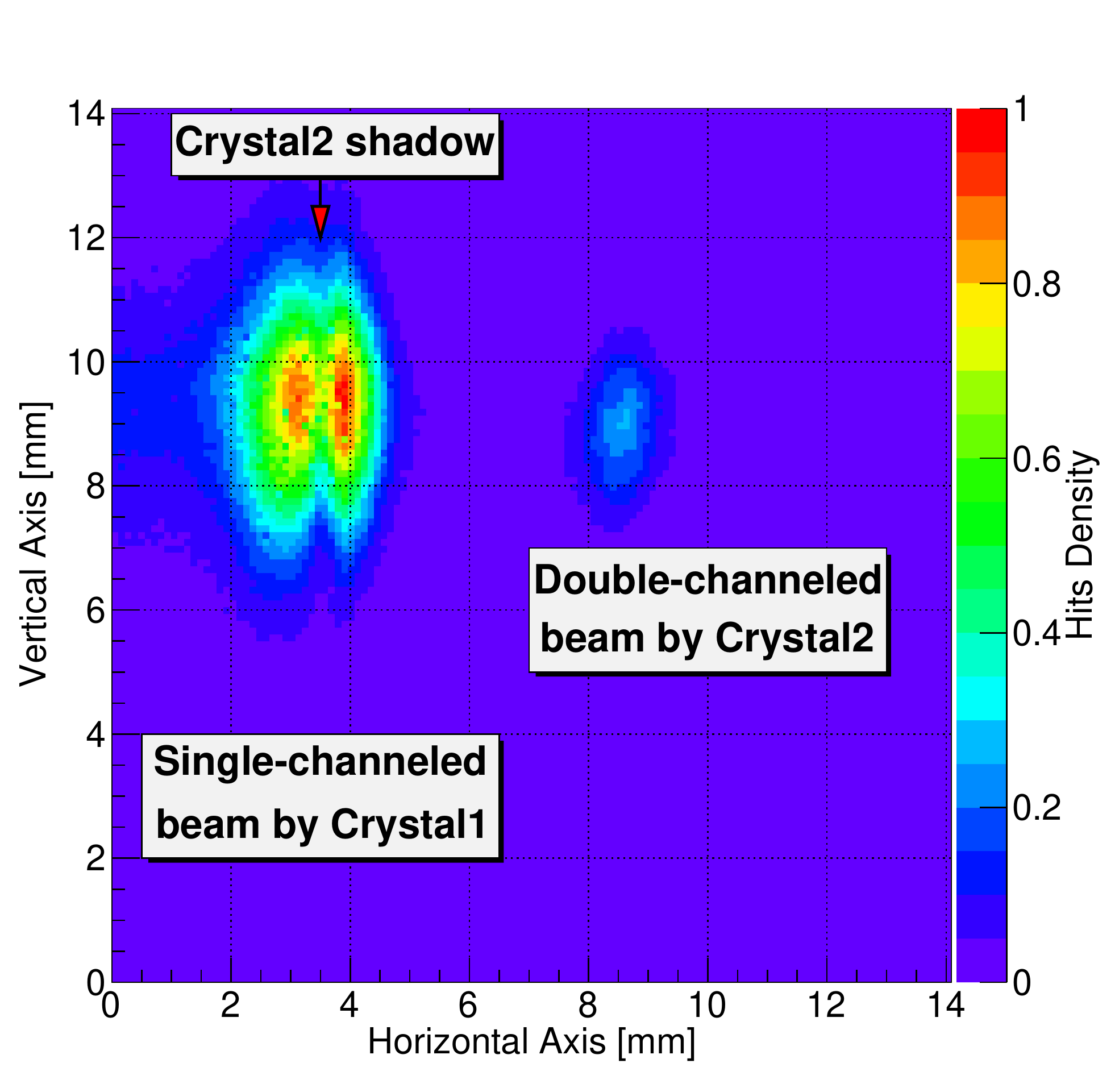}
    \captionof{figure}{(Color online.) Timepix image of the double-channeled beam spot.}
\label{fig:fig5}
\end{minipage}

\begin{minipage}{0.9\linewidth}
    \centering\includegraphics[width=1\linewidth]{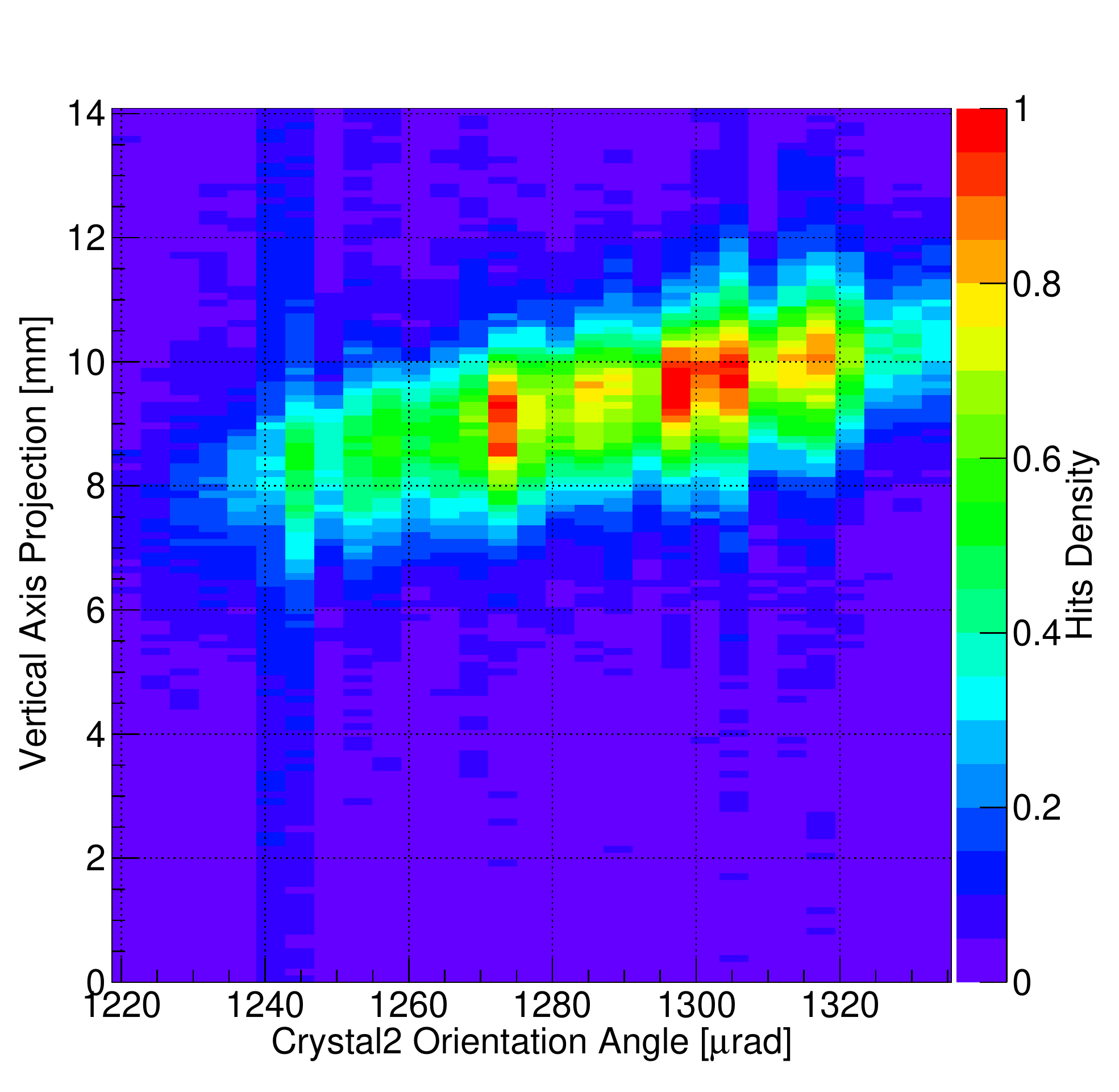}
    \captionof{figure}{(Color online.) Timepix image projection on a vertical axis as a function of the Crystal2 orientation angle. The projection has been done only for the double-channeled beam spot.}
\label{fig:fig6}
\end{minipage}

%\begin{figure}[htbp]
%\centering
%\includegraphics[width=1\textwidth]{Figure258}
%\caption{Timepix image of the double-channeled beam spot.}
%\label{fig:fig5}
%\end{figure}

%\begin{figure}[htbp]
%\centering
%\includegraphics[width=1\textwidth]{Figure251}
%\caption{Timepix image projection on a vertical axis as a function of the Crystal2 orientation angle. The projection has been done only for the double-channeled beam spot.}
%\label{fig:fig6}
%\end{figure}

Such a configuration of crystals was used for the first time at the circulating machine for \textit{in~situ} measurements of the bent crystal twisting, called torsion (different atomic planes orientations along the crystal's height). The measured value of the torsion is equal to 73.7$\pm$1.4~$\mu$rad/mm. Figure~\ref{fig:fig6} shows a vertical movement of the double-channeled beam spot during Crystal2 angular scan placed inside a single-deflected beam.

\section{Double-channeling measurements}

\begin{minipage}{0.9\linewidth}
    \centering\includegraphics[width=1\linewidth]{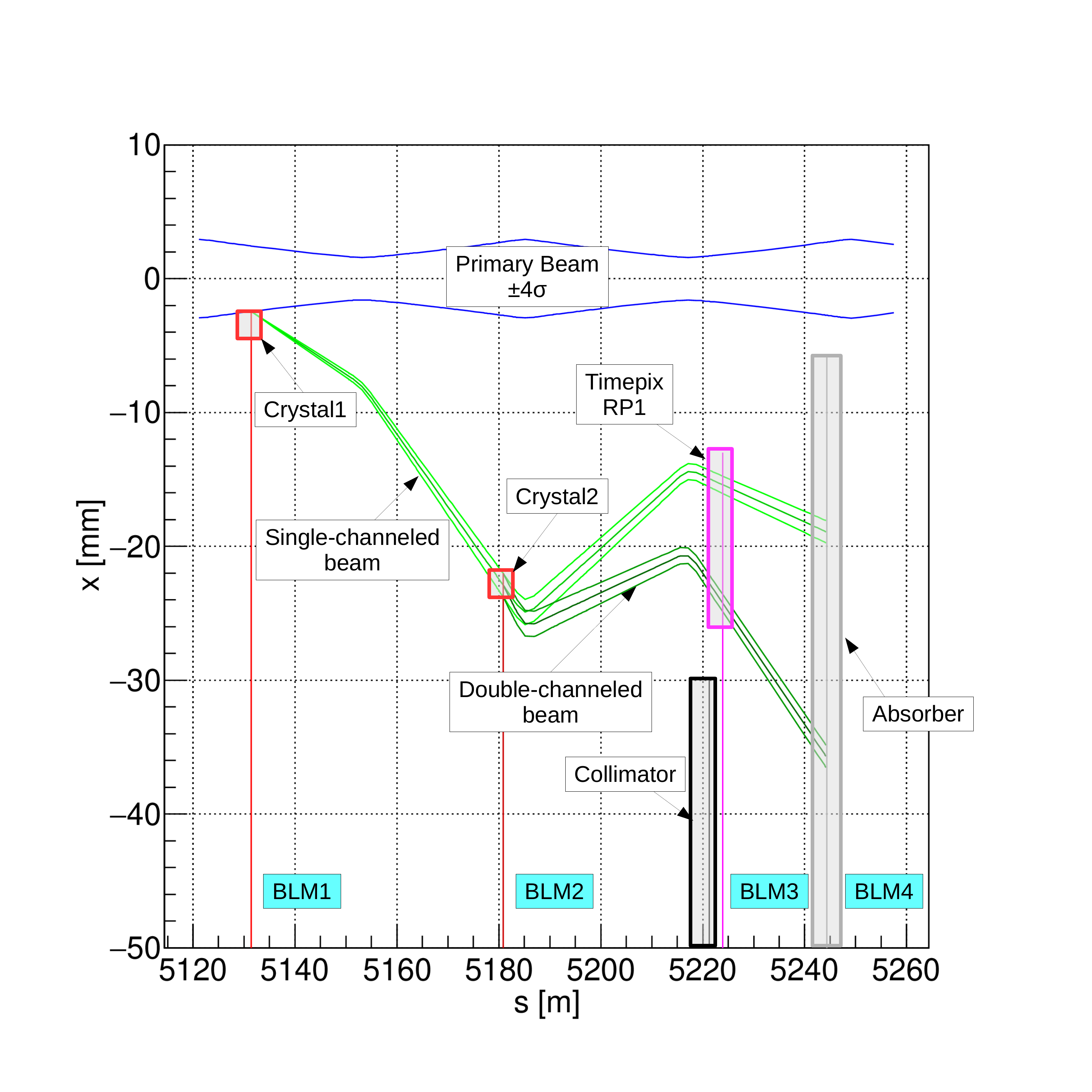}
    \captionof{figure}{(Color online.) Horizontal projection ($x$) of the channeled halo particles as a function of the longitudinal coordinate ($s$) in LSS5 zone of the CERN SPS. Blue lines represent the primary beam envelope with a 4$\sigma$ width. An upstream Crystal1 (red line) deflects halo particles (green line), within $\pm \theta_{\rm c}$ range (light green lines) onto the tungsten absorber (grey line), as well as channeled particles by the downstream Crystal2 (dark red line), inserted into the deflected beam by the upstream crystal with a phase advance of about 60$^\circ$, to maximize the separation between channeled and primary beams. Timepix (pink line) detector, located between the crystal and absorber, intercepts the channeled particles, the collimator jaw (black line) is retracted from the beam. Each component of the setup is followed by the respective BLM counter, located outside the beam pipe.}
\label{fig:fig1}
\end{minipage}

The main limitation factor of the test in 2017 was a small cross-section (0.5~mm) of the available crystal in the downstream position. Therefore, during the 2017--2018 Winter Technical Stop a set of new crystals was installed (Tab.~\ref{tab:tab1}, from the PNPI Gatchina). Table~\ref{tab:tab2} shows the SPS beam conditions for the measurements, while particle trajectories are shown in Figure~\ref{fig:fig1}.

\begin{minipage}{0.9\linewidth}
\centering
\captionof{table}{Crystal parameters, where $\theta_{\rm def}^{\rm H8}$ is a deflection angle of the crystal measured at the H8 beamline, $L$ is its length along the beam, while $W$ is a width of the crystal transversely to the beam direction. $AC$ means an anti-clastic crystal bending method.}\label{tab:tab1} 
\begin{tabular}{c|c|c|c|c}
	&Material&$L$&$W$&$\theta_{\rm def}^{\rm H8}$\\
	&(Plane)&[mm]&[mm]&[$\mu$rad]\\
	&&($\pm$0.02)&($\pm$0.02)&($\pm$1)\\
    \hline
    Crystal1 &Si(110)$AC$&4.0&1.5&301.0\\    
    Crystal2 &Si(110)$AC$&6.0&4.0&196.8\\
\end{tabular}
\end{minipage}

\begin{minipage}{0.9\linewidth}
\centering
\captionof{table}{SPS beam conditions during the measurements. A single 3~ns long bunch circulates inside the machine with a frequency of 43~kHz.}\label{tab:tab2} 
\begin{tabular}{c|c|c|c|c}
	Beam&Momentum&Intensity&Emittance&Tune\\
	& [GeV/c]&per bunch&[nm$\cdot$rad]&$Q_{\rm h}$\\
	\hline
	proton&270&$\rm\sim 10^{11}$&$\sim$5&20.13\\
\end{tabular}
\end{minipage}

The optimal position of the downstream crystal with respect to the single-deflected particle beam was provided by a dedicated linear scan. Crystal2 in the amorphous orientation creates an obscurity of particle counts on the Timepix image horizontal projection intercepting a single-channeled beam, which is illustrated in Figure~\ref{fig:fig2}. The measured shadow width of $W_{\rm sh} = 4.15 \pm 0.14$~mm is in a good agreement with a nominal transverse size of Crystal2 $W = 4.00 \pm 0.02$~mm (Tab.~\ref{tab:tab1}).

\begin{minipage}{0.9\linewidth}
    \centering\includegraphics[width=1\linewidth]{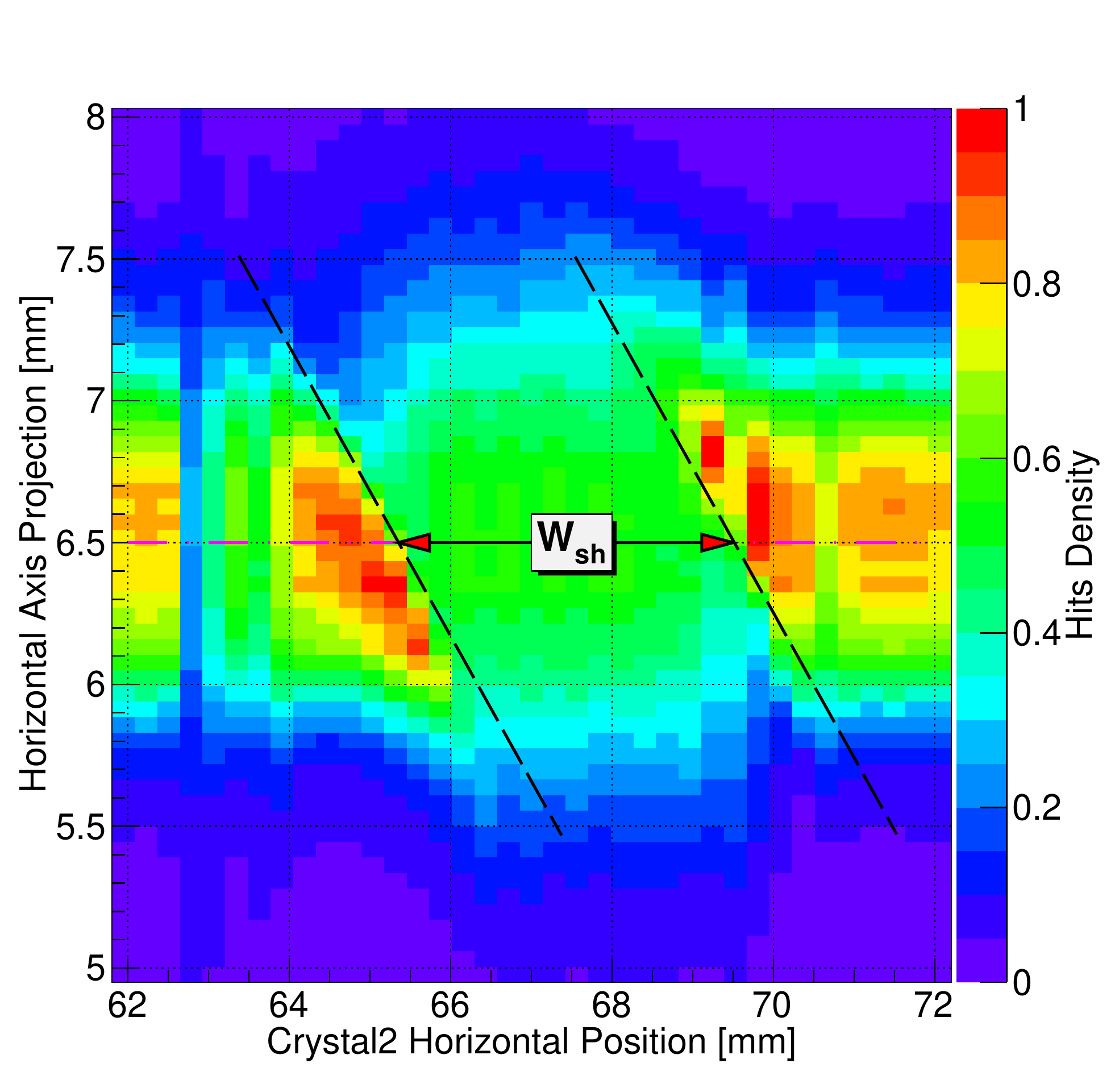}
    \captionof{figure}{(Color online.) Timepix horizontal projection for the Crystal2 optimal position. The width ($W_{\rm sh}$) of the shadow is measured between two edges of the particle scattered regions (black dashed lines).}
\label{fig:fig2}
\end{minipage}

\begin{minipage}{0.9\linewidth}
    \centering\includegraphics[width=1\linewidth]{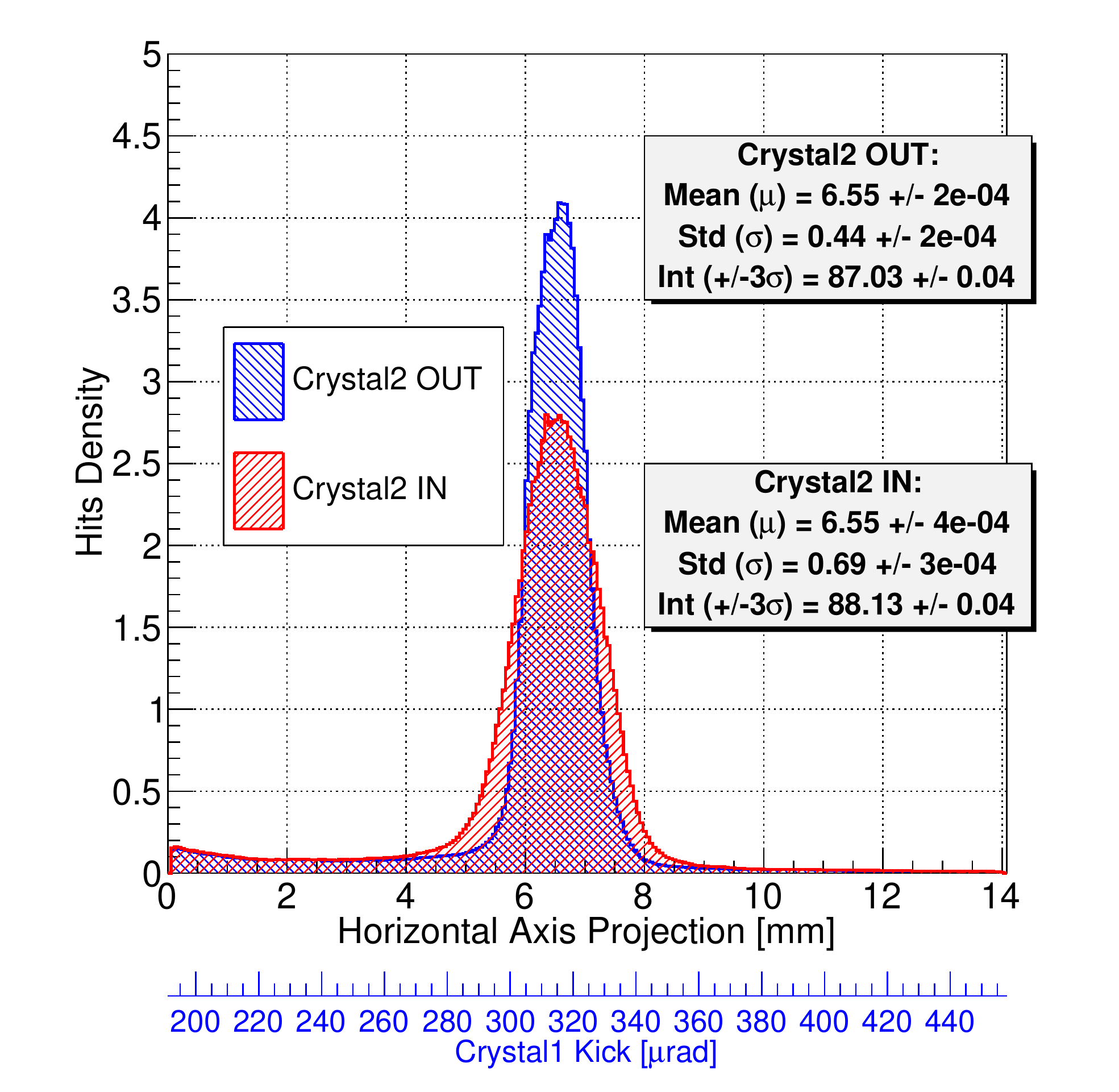}
    \captionof{figure}{(Color online.) Horizontal single-channeled beam profile measured by means of the Timepix detector at RP1. Two distributions correspond to the following cases: the downstream crystal is retracted from a single-deflected beam ("Crystal2 OUT"), and the crystal is perfectly aligned at the center of the beam ("Crystal2 IN").}
\label{fig:fig3}
\end{minipage}

Figure~\ref{fig:fig3} shows two horizontal profiles of the single-channeled beam, when the downstream crystal was retracted from the beam ("Crystal2 OUT", horizontal position of 60-61~mm), and when it was perfectly aligned with respect to the center of the beam ("Crystal2 IN", horizontal position of 67-68~mm). For Crystal2~IN case, it can be seen an increase of the particle spread ($\sigma$) compared to Crystal2~OUT situation. However, due to the constant particle flux, the integrals ($Int$) of the counts within $\pm3\sigma$ for two distributions are almost equal (about 1\% of difference). In turn, this means that the inelastic nuclear interaction contribution is negligible.

The angular distribution characteristics are listed in Table~\ref{tab:tab3}. The difference between the two cases are induced by the multiple Coulomb scattering (MCS) of protons inside the downstream crystal. According to~\cite{MDM1} the angular dispersion $\sigma_{\rm MCS}$ due to the MCS for a crystal, which contributes 6~mm of silicon, is about 11~$\mu$rad for a 270~GeV/c proton. Therefore, this value is in a good agreement with our measurements: 
\begin{equation}
\sqrt{\sigma_{\rm Crystal2~IN}^{2} - \sigma_{\rm Crystal2~OUT}^{2}} = 10.02\pm0.01\rm~\mu rad.
\end{equation}
Taking into account the beta function values ($\beta^{\rm CR2}_{\rm x}=93.5875$~m, $\beta^{\rm TPX}_{\rm x}=40.1684$~m), the standard deviation of the single-channeled beam in the Crystal2 reference frame can be computed as following: 
\begin{equation}
\sigma^{\rm CR2}_{\rm x} = \sigma^{\rm TPX}_{\rm x} \sqrt{{\beta^{\rm CR2}_{\rm x} \over \beta^{\rm TPX}_{\rm x}}} = 0.67\pm3\cdot 10^{-4}\rm~mm,
\end{equation}
where $\sigma^{\rm CR2}_{\rm x}$ and $\sigma^{\rm TPX}_{\rm x}$ are the standard deviations of the beam at Crystal2 and Timepix locations respectively. Therefore, in a perfect aligned position of the downstream crystal, it covers almost entire single-channeled beam (within $\pm 3 \sigma = 4.03 \pm 2\cdot 10^{-3}$~mm).

\begin{minipage}{0.9\linewidth}
\centering
\captionof{table}{Crystal1 kick angle characteristics, where $\mu$ is a mean value, while $\sigma$ is a standard deviation.}\label{tab:tab3} 
\begin{tabular}{c|c|c}
	Configuration&$\mu$ [$\mu$rad]&$\sigma$ [$\mu$rad]\\
    	\hline
    "Crystal2 OUT"&315.34$\pm$4e-3&8.44$\pm$5e-3\\
    "Crystal2 IN"&315.25$\pm$7e-3&13.10$\pm$9e-3\\
\end{tabular}
\end{minipage}

\begin{minipage}{0.9\linewidth}
    \centering\includegraphics[width=1\linewidth]{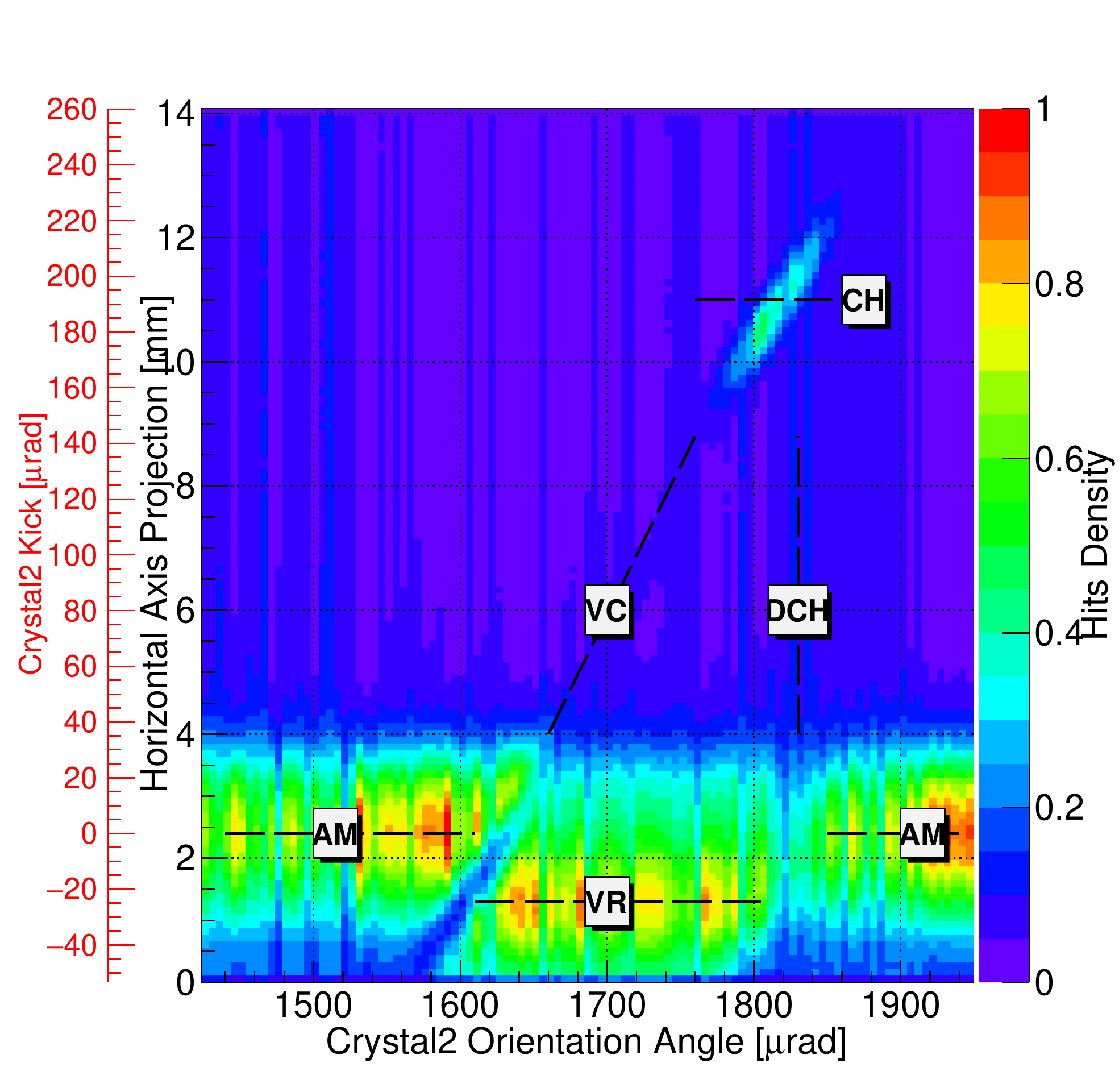}
    \captionof{figure}{(Color online.) Timepix image projection on the horizontal axis as a function of the Crystal2 orientation angle, for 1~$\mu$rad/s of the goniometer rotational speed. Timepix detector was operated in the Medipix mode with a 48~MHz clock, and 0.5~s of the acquisition time window. Each Timepix image projection is normalized by the beam intensity. Black dashed lines represent different regions of the deflection angle distribution. Bin size is 0.11~mm$\times$5~$\mu$rad.}
\label{fig:fig4}
\end{minipage}

An angular scan of Crystal2 (Fig.~\ref{fig:fig4}), placed inside a single-deflected beam, illustrates a typical particles distribution seen earlier at the H8 beamline~\cite{CHANNELING_14}. It starts from the amorphous orientation (AM) of the crystal. At around 1600~$\mu$rad the volume reflection (VR) effect starts to dominate during about 200~$\mu$rad of the scan, which is close to the bending angle of the crystal (Tab.~\ref{tab:tab1}). The VR process is accompanied by the volume capture (VC), while at the perfect channeling orientation (CH) a deflected beam spot is located at around 1800~$\mu$rad with a low populated dechanneling region (DCH).

Figure~\ref{fig:fig7} shows an integrated 2D image of the double-deflected beam, measured by means of the Timepix detector. A not perfect crystal orientation for the particle channeling results in the asymmetrical shape of the non-channeled beam (the left beam spot on the figure) due to the volume reflection process contribution.

\begin{minipage}{0.9\linewidth}
    \centering\includegraphics[width=1\linewidth]{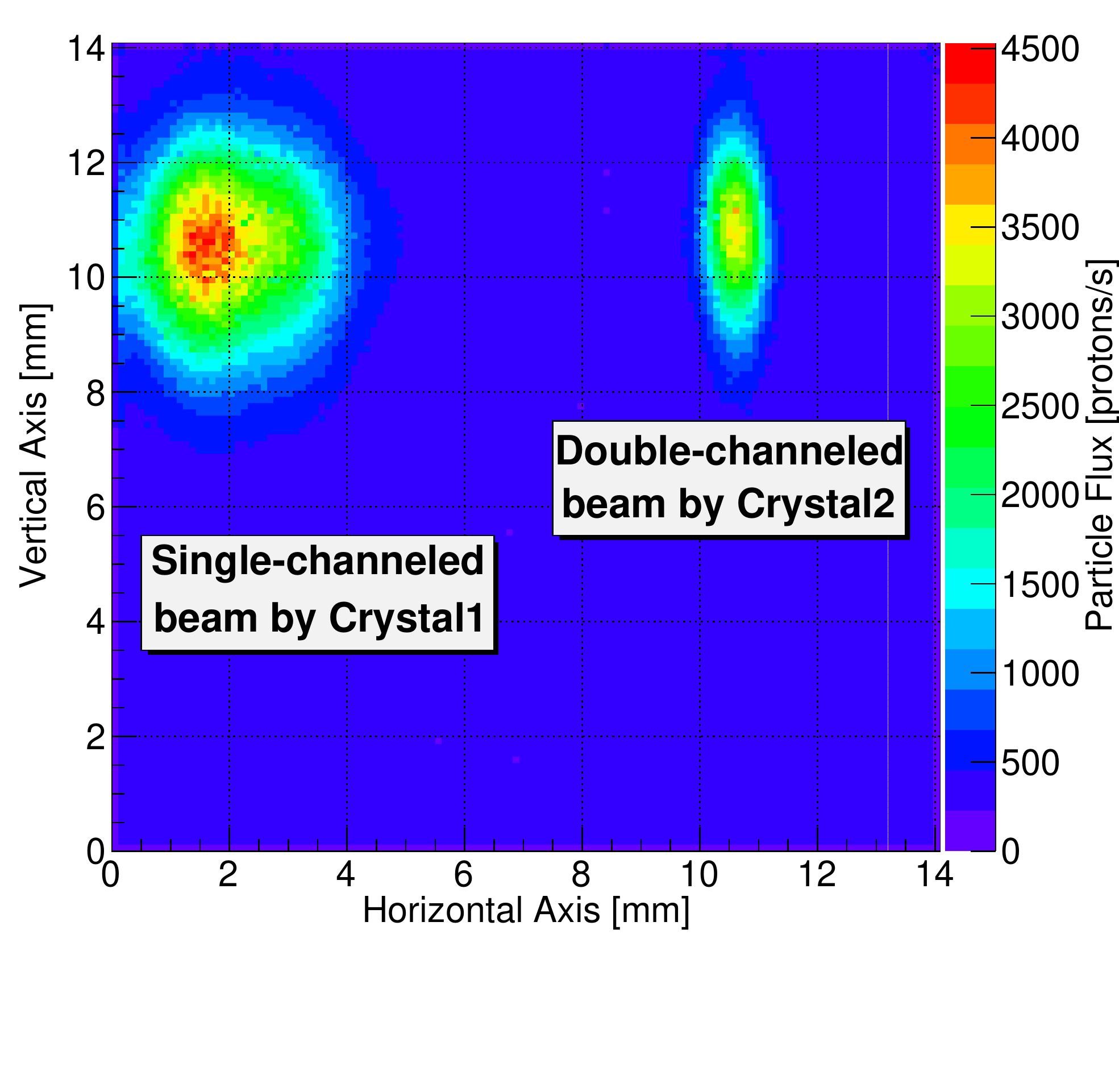}
    \captionof{figure}{(Color online.) Timepix integrated image of the double-channeling. Medipix mode, 0.5 s of the acquisition time window with 48~MHz clock, integration over 175~frames. Bin size is 0.11$\times$0.11~mm$^{2}$.}
\label{fig:fig7}
\end{minipage}

\begin{minipage}{0.9\linewidth}
    \centering\includegraphics[width=1\linewidth]{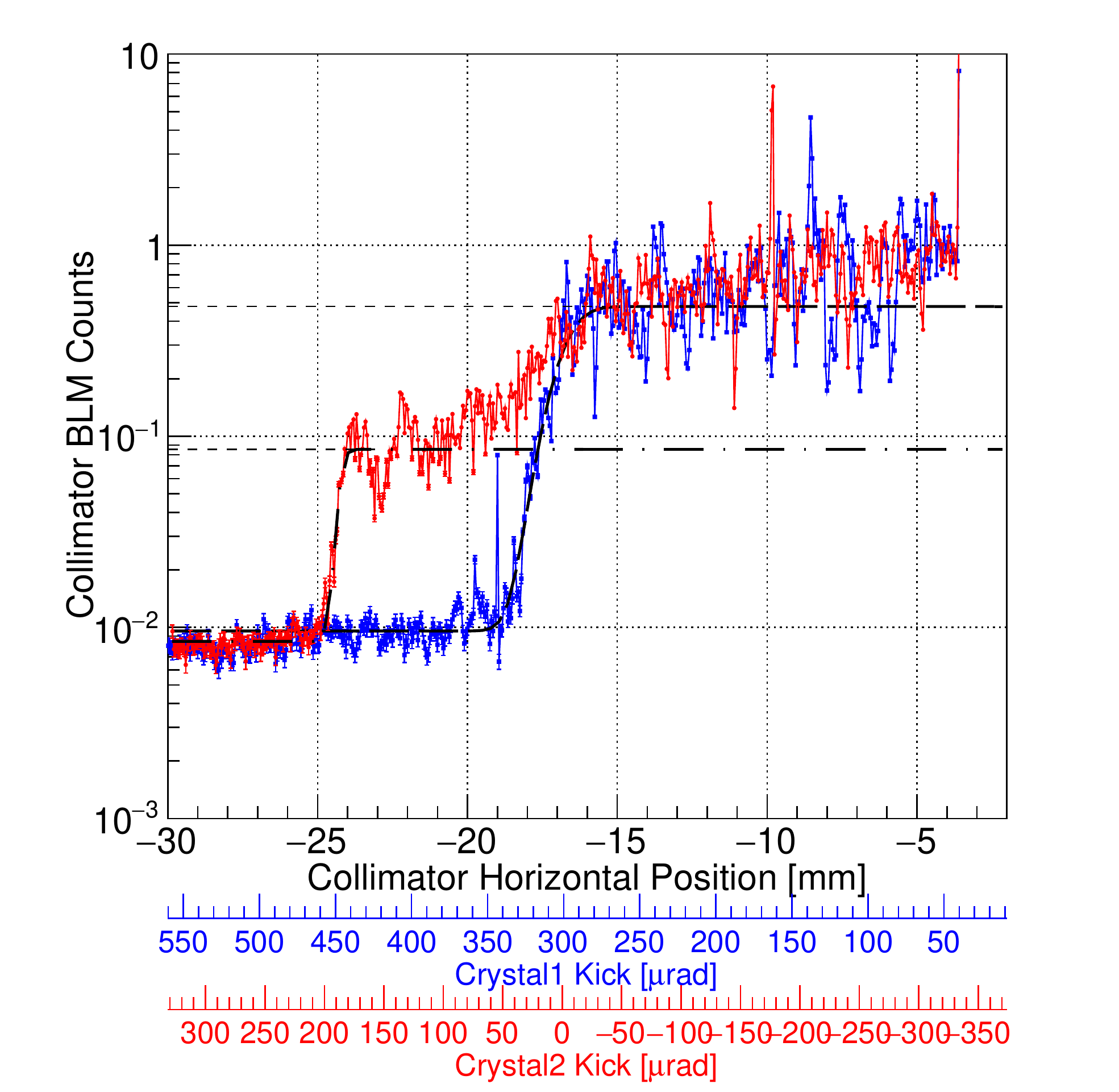}
    \captionof{figure}{(Color online.) BLM signal for the collimator jaw linear scan across single- and double-channeled beams. Linear speed of 50~$\mu$m/s. The spike at around -6~mm is produced by the jaw touching the primary beam. Black dashed and dot-dashed lines illustrate a fit of the distributions by means of the Error Functions for single- and double-channeled beams respectively. BLM counts are normalized to the beam intensity.}
\label{fig:fig8}
\end{minipage}

In such a configuration, the Timepix measurements are compared with two linear scans of the collimator jaw. A blue line in Figure~\ref{fig:fig8} introduces local beam losses induced by the collimator crossing only a single-channeled beam, and Crystal2 is in AM. For a double-channeling configuration, a red line follows the losses in Figure~\ref{fig:fig8} when the jaw intercepts double-, and then single-deflected beams. According to~\cite{CHANNELING_22}, the value of BLM counts at the moment when the jaw touches the main SPS beam (i.e. it becomes the primary obstacle on the beam) is proportional to the impinging particle flux onto the upstream Crystal1. Therefore single- and double-channeling plateaus from the Error function fit correspond to the deflection efficiency with respect to the impinging particle flux. The calculated value of the multiturn channeling efficiency for Crystal1 is $\epsilon^{\rm CR1}_{\rm MtCH} = 0.478 \pm 0.027$. However, the double-channling deflection efficiency of the two crystals setup is equal to $\epsilon_{\rm DCH} = 0.085 \pm 0.004$. In such a way, a computed single-pass deflection efficiency from the BLM measurements is: $\epsilon^{\rm CR2}_{\rm SpCH} = \epsilon_{\rm DCH}/\epsilon^{\rm CR1}_{\rm MtCH} = 0.179 \pm 0.013$, while from the Timepix this value is equal to $0.188 \pm 3 \cdot 10^{-5}$.

\section{Conclusions}

Examining the obtained results, we can conclude that the developed methodology of the double-crystal setup alignment has been successfully verified for the high-energetic particle accelerator. Measured deflected particle beam parameters are in a good agreement with our expectations. A quantum imaging pixel detector Timepix provides quantitative measurements of the beam profile parameters with a possibility to identify various bent crystal operational regimes (e.g. particle channeling, volume reflection, etc.), and also to measure a single-pass channeling efficiency. The deflection ability of the two crystals can be improved by the optimization of their parameters. For the first time at the circulating machine, a double-crystal setup has been used to quantify a downstream crystal torsion value. 

Following the PBC project program for the fixed target experiment, further double-crystal investigations of the particles channeling and induced background by a tungsten target installed in front of the downstream crystal are ongoing.

\section*{Acknowledgements}

The authors are grateful to other members of the UA9 Collaboration and CERN EN/STI group for their support and contribution to the development of the experimental apparatus used in this study.

\bibliography{main}

\end{multicols}
\end{document}